\documentclass[iop]{emulateapj}

\slugcomment{Received 2016 January 21; accepted 2016 February 17; will
be published in AJ}

\usepackage{ulem}
\usepackage{url}
\usepackage{color}

\shorttitle{S208}

\shortauthors{Yasui et al.}

\begin{document}

\title{Low-metallicity Young Clusters in the Outer
Galaxy. II. S\lowercase{h} 2-208}

\author{Chikako Yasui\altaffilmark{1, 2}, Naoto
Kobayashi\altaffilmark{3, 4, 2}, Masao Saito\altaffilmark{5, 6}, and
Natsuko Izumi\altaffilmark{3, 2}}

\affiliation{$^1$Department of Astronomy, Graduate School of Science,
University of Tokyo, Bunkyo-ku, Tokyo 113-0033, Japan;
{ck.yasui@astron.s.u-tokyo.ac.jp} \\
$^2$Laboratory of Infrared High-resolution spectroscopy (LIH), Koyama
Astronomical Observatory, Kyoto Sangyo University, Motoyama, Kamigamo,
Kita-ku, Kyoto 603-8555, Japan \\
$^3$Institute of Astronomy, School of Science, University of Tokyo,
2-21-1 Osawa, Mitaka, Tokyo 181-0015, Japan \\
$^4$Kiso Observatory, Institute of Astronomy, School of Science,
University of Tokyo, 10762-30 Mitake, Kiso-machi, Kiso-gun, Nagano
397-0101, Japan \\
$^5$Nobeyama Radio Observatory, 462-2 Nobeyama,
Minamimaki-mura, Minamisaku-gun, Nagano 384-1305, Japan \\
$^6$The Graduate University of Advanced Studies,
(SOKENDAI), 2-21-1 Osawa, Mitaka, Tokyo 181-8588, Japan}


%
\begin{abstract}
We obtained deep near-infrared images of Sh 2-208, one of the
lowest-metallicity \ion{H}{2} regions in the Galaxy, ${\rm [O/H]} =
-0.8$\,dex. 
We detected a young cluster in the center of the \ion{H}{2} region with
a limiting magnitude of $K=18.0$\,mag (10$\sigma$), which corresponds to
a mass detection limit of $\sim$0.2\,$M_\odot$.
This enables the comparison of star-forming properties under low
 metallicity with those of the solar neighborhood.
We identified 89 cluster members. 
From the fitting of the {\it K}-band luminosity function (KLF), the age
and distance of the cluster are estimated to be $\sim$0.5\,Myr and
$\sim$4\,kpc, respectively.
The estimated young age is consistent with the detection of strong CO
emission in the cluster region and the estimated large extinction of
cluster members ($A_V \sim 4$--25\,mag).
The observed KLF suggests that the underlying initial mass function
(IMF) of the low-metallicity cluster is not significantly different from
canonical IMFs in the solar neighborhood in terms of both high-mass 
slope and IMF peak (characteristic mass).
Despite the very young age, the disk fraction of the cluster is
estimated at only 27$\pm$6\,\%, which is significantly lower than those
in the solar metallicity.
Those results are similar to Sh 2-207, which is another low-metallicity
star-forming region close to Sh 2-208 with a separation of 12\,pc,
suggesting that their star-forming activities in low-metallicity
environments are essentially identical to those in the solar
neighborhood, except for the disk dispersal timescale.
From large-scale mid-infrared images, we suggest that sequential star
formation is taking place in Sh 2-207, Sh 2-208 and the surrounding
region, triggered by an expanding bubble with a $\sim$30\,pc radius.

\end{abstract}





\section{INTRODUCTION} \label{sec:intro}

The study of young low-metallicity clusters is of great interest because
they enable us to characterize the metallicity dependence of star
formation processes, such as the initial mass function (IMF) and
star formation efficiency.
Also, the study of the disk fraction for young clusters can constrain
the probability of planet formation.
They also provide an opportunity to study star formation in the early
universe and to understand the star formation process itself
\citep{Yasui2016}.
Although nearby dwarf galaxies, such as the Large Megallanic Cloud (LMC)
and the Small Megallanic Cloud (SMC), have been the primary targets of 
such studies (e.g., \citealt{Brandl1996}), our Galaxy also
harbors a variety of low-metallicity clusters that enable the 
study of such objects with unprecedented detail.

The thin disk of our Galaxy is known to have a metallicity gradient,
lower in the outer Galaxy, as well as other spiral galaxies
\citep{Rudolph2006}.
Although the existence of the metallicity gradient in the inner Galaxy
($R_G \lesssim 8$\,kpc) is still under debate, a trend toward low
metallicity can generally be seen in the outer Galaxy \citep{Bono2013}.
For the derivation of the metallicity, various kinds of objects are
used, including B-type stars (e.g., \citealt{Smartt2001}), Cepheid
\citep[e.g.][]{Luck2006}, \ion{H}{2} regions
\citep[e.g.][]{Rudolph2006}, and open clusters (e.g.,
\citealt{Yong2012}).
Among these objects, \ion{H}{2} regions have been found over a wide
range of Galactic radii beginning long ago through a variety of
H$\alpha$ emission surveys, starting with the pioneering work by
\citet{Sharpless1953}.
\citet{Sharpless1959} compiled a catalog of 313 \ion{H}{2} regions that
is comprehensive for north of decl. $-$27$^\circ$, covering the galactic
longitudes of 0--272\,deg and 315--360\,deg.
The Galactrocentric distance of the regions span from $\sim$0 to 18\,kpc
\citep{{Russeil2003},{Foster2015}}.
Considering that \ion{H}{2} regions are excited by O- or B-type stars
with very short lifetimes of $\sim$20\,Myr at most, a large number of
them are star-forming regions ($\sim$70\,\%, \citealt{Blitz1982}).
Therefore, we selected seven \ion{H}{2} regions with estimated low
metallicity, [O/H$] \le -0.5$\,dex, in the literature to construct a
sample of low-metallicity young clusters.

In a series of papers, we present observations regarding the properties
of young, low-metallicity clusters.
In the first paper \citep{Yasui2016}, we presented the results for Sh
2-207 (S207), which is an \ion{H}{2} region with a $\sim$2.5\,arcmin
radius in the second Galactic quadrant with a metallicity of
$-$0.8\,dex.
In the western region of S207, we identified a young cluster with a
$\sim$1\,arcmin radius (corresponding to 1.3\,pc) and a limiting
magnitude of $K_S = 19.0$\,mag (10$\sigma$) that corresponds to a mass
detection limit of $\lesssim$0.1\,$M_\odot$.
The age and distance of the cluster are estimated at 2--3\,Myr and
$\sim$4\,kpc, respectively.
The observed $K$-band luminosity function (KLF) suggests that the
underlying IMF of the cluster down to the detection limit is not
significantly different from typical IMFs in the solar metallicity.
Based on the fraction of stars with near-infrared (NIR) excesses, a low
disk fraction ($<$10\,\%) in the cluster with the relatively young age
is suggested \citep[see also][]{{Yasui2010}}.

In this paper, we present the results for the second target, Sh 2-208
(S208), which is one of the lowest-metallicity star-forming regions in
the Galaxy with $12+\log ({\rm O/H}) \le 8$ \citep{Rudolph2006}.
S208 is located $\sim$10\,arcmin southward of S207. 
The two clusters consist of an ideal pair for studying low-metallicity
clusters because their proximity in on-sky position and identical
metallicities suggest that they are in similar environments.
The combination of the S207 and S208 clusters may enable a unique study
of the evolution of IMF and other star formation parameters (e.g., disk
fraction, which is strongly related to the formation rate of the
protoplanetary disk) in low-metallicity environments.

This paper is organized as follows. Section~2 describes previous studies
on S208 and its star-forming activities using
{\it WISE (Wide-field Infrared Survey Explorer)} and
2MASS (Two Micron All Sky Survey) data. 
Section~3 describes Subaru MOIRCS (Multi-Object InfraRed Camera and
Spectrograph) deep $JHK_S$ images and data reduction. 
Section~4 describes the results for a star-forming cluster in S208. In
Section~5, we discuss the implications for the basic cluster parameters,
such as age, distance, IMF, and disk fraction.
Finally, in Section~\ref{sec:SFinOA}, we discuss the formation of those
clusters in the outer arm in the Galactic context.

\section{S\lowercase{h} 2-208} \label{sec:S208} 

\begin{figure*}[t]
\begin{center}
\includegraphics[width=13cm]{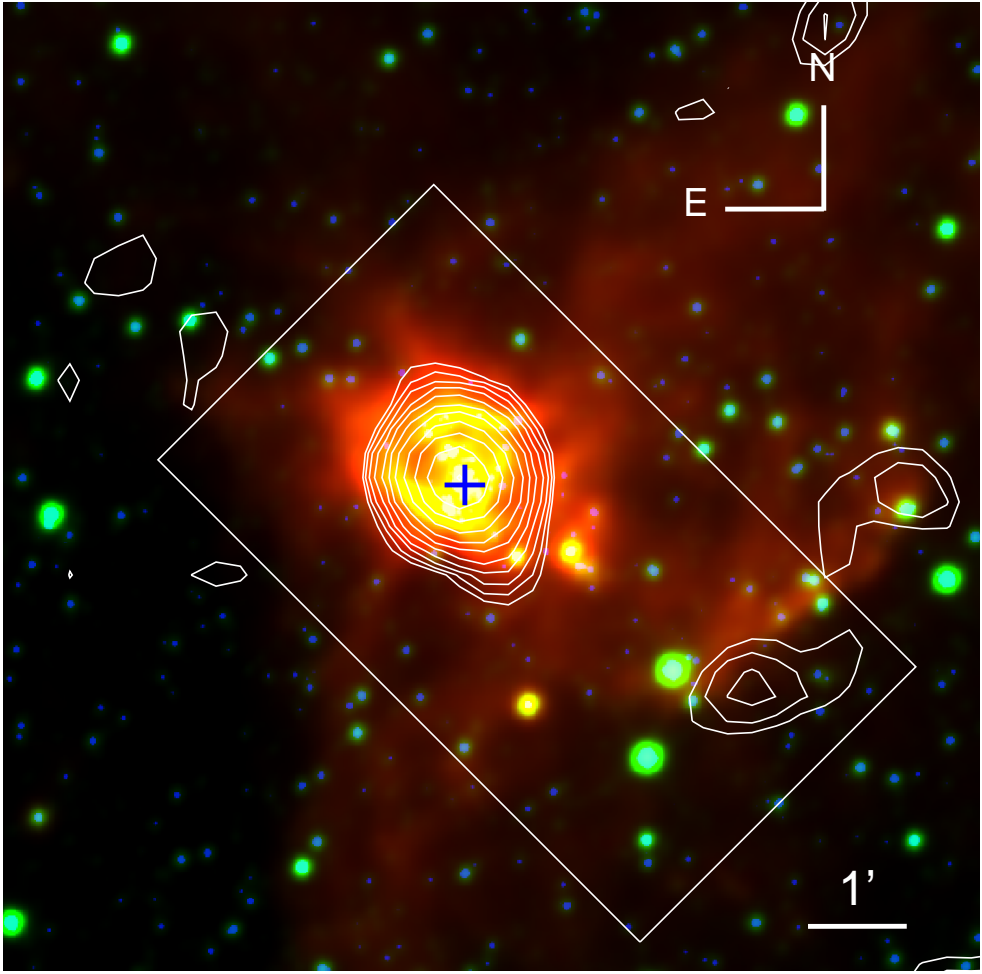}
\caption{Pseudocolor image of S208 with 
the center of $(\alpha_{\rm 2000.0}, \delta_{\rm 2000.0}) = (04^{\rm h}
19^{\rm m} 32.92^{\rm s}, +52^\circ 58' 41.6'')$ in Equatorial
coordinates and $(l, b) = (151.2870^\circ, +1.9682^\circ$) in Galactic
coordinates.
The size of the field is 10$'\times$10$'$.  North is up and east is to
the left. 
The 1\,arcmin corresponds to 2.7\,pc and 1.2\,pc for distances of S208 
of 9\,kpc and 4\,kpc, respectively.
The image is produced by combining the 2MASS $K_S$-band (2.16\,$\mu$m;
blue), {\it WISE} band 1 (3.4\,$\mu$m; green), and {\it WISE} band 3
(12\,$\mu$m; red).
The 1.4\,GHz radio continuum emission by NVSS is also shown using the
 white contours.  The contours are plotted at 1\,mJy\,${\rm beam}^{-1} 
 \times 2^0, 2^{-1/2}, 2^1$, .... 
The blue plus symbol shows the brightest star in the optical bands (GSC
 03719-00517). 
The white box shows the location and size of the MOIRCS field of
 view. }
\label{fig:3col_2MASS_WISE} 
\end{center}
\end{figure*}

In this section, the properties of the target star-forming region, S208,
are summarized.
In Table~\ref{tab:targets}, we summarize the properties from previous
works: the coordinate, distance, oxygen abundance, and metallicity.
We also show NIR and mid-infrared (MIR) pseudocolor images of S208 in
Figure~\ref{fig:3col_2MASS_WISE}.


\subsection{Basic Properties from the Literature} \label{sec:properties}

\setcounter{footnote}{6}

S208 is located at $(l, b) = (151.2870^\circ, +1.9682^\circ)$ on the
Galactic plane with coordinates of $(\alpha_{\rm 2000.0}, \delta_{\rm 
2000.0}) = (04^{\rm h} 19^{\rm m} 32.92^{\rm s}, +52^\circ 58' 41.6'')$
from SIMBAD\footnote{This research has made use of the SIMBAD database,
operated at the Centre de Donn\'ees Astronomiques de Strasbourg,
France.} \citep{Wenger2000}.
It has an extended \ion{H}{2} region traced by H$\alpha$ emission
\citep{Sharpless1959} and radio continuum emission
\citep{Fich1993}. 
It also accompanies strong MIR emission that is detected with IRAS (IRAS
04156+5251 in the IRAS Point Source Catalog; \citealt{Beichman1988}). 
CO emission is reported in \citet{Blitz1982} and \citet{Wouterloot1989}.
A star-forming cluster is identified by \citet{Bica2003} using 2MASS
images, with the center of $(\alpha_{\rm 2000.0}, \delta_{\rm 2000.0}) =
(04^{\rm h} 19^{\rm m} 33^{\rm s}, +58^\circ 52' 42'')$ and the angular
dimensions of 1.2$'$.
The photometric distance, which is determined from spectroscopic and
photometric observations, is estimated to be $\simeq$9\,kpc on average
for a probable dominantly exciting late O-type or early B-type
star\footnote{The spectral type is estimated at B0V by 
\citet{Crampton1978} and \citet{Moffat1979}, while it is estimated to be
O9.5V by \citet{Chini1984} and \citet{Lahulla1985}.}, GSC 03719-00517
(7.6\,kpc by \citealt{Moffat1979}, 9.4\,kpc by \citealt{Chini1984},
10.0\,kpc by \citealt{Lahulla1985}; see the blue plus in
Figure~\ref{fig:3col_2MASS_WISE}).
Assuming that the Galactocentric distance of the Sun is $R_\odot =
8.0$\,kpc, then the distance corresponds to $R_G \simeq 16.5$\,kpc. 
It should be noted that the distance from such early-type stars may not
be accurate because the luminosity class cannot be easily identified
for early-type stars. 
On the other hand, the kinematic distance of S208 is estimated at about
4\,kpc based on observations of the H$\alpha$ radial velocity using a
Fabry--Perot spectrometer
(4.0\,kpc by \citealt{Wouterloot1989}, 4.1\,kpc by
\citealt{Caplan2000}), and from radial velocities of \ion{H}{1} data by
CGPS and $^{12}$CO line data by FCRAO (4.4\,kpc by
\citealt{Foster2015}).
In this case, the distance corresponds to $R_G \simeq 12$\,kpc.
Figure~\ref{fig:Gal} shows a top view of the Galaxy with S208 and the
spiral arms. 
The locations of S208, assuming the photometric distance and kinematic
distance, are shown with the open circle and the filled circle,
respectively, while the spiral arms are shown with different colors,
e.g., red for the Norma--Cygnus (outer) arm.
The figure shows that S208 is located beyond the outer arm in the case
of photometric distance, whereas it is located around the outer arm in
the case of the kinematic distance. 
Later, we will propose that the distance to this cluster is more likely
4\,kpc based on KLF analysis, which is similar to the above kinematic 
distance.

Based on the Fabry--Perot observations, \citet{Caplan2000} measured
several optical emission line fluxes in [\ion{O}{2}]
$\lambda\lambda$3726 and 3729, H$\beta$, [\ion{O}{3}] $\lambda$5007,
[\ion{He}{1}] $\lambda$5876, and H$\alpha$, for 36 \ion{H}{2} regions
including S208.
Subsequently, \citet{Deharveng2000} derived the oxygen abundance (O/H),
as well as the extinctions, electron densities and temperatures, and
ionic abundances (O$^+$/H$^+$, O$^{++}$/H$^+$, and He$^+$/H$^+$). 
The estimated oxygen abundance of S208 is $12 + \log 
({\rm O/H}) = 8.00$. 
\citet{Rudolph2006} reanalyzed the elemental abundances of 117
\ion{H}{2} regions with updated physical parameters.
Among the \ion{H}{2} regions, the oxygen abundance of S208 is estimated
using the data of \citet{Caplan2000} to be $12 + \log ({\rm O/H}) =
7.91^{+0.16}_{-0.26}$.
This corresponds to the metallicity of ${\rm [O/H]} \simeq -0.8$\,dex
assuming the solar abundance of $12 + \log ({\rm O/H}) = 8.73$
\citep{Asplund2009}.
The spatial distribution of the Galactic abundance using the
spectroscopy of Cepheids \citep{Luck2006} also suggests low metallicity
($\lesssim$$-$0.5\,dex) in the outer Galaxy at a distance of $D \gtrsim
4$\,kpc in the second quadrant, where S208 is located,
although it should be noted that some recent studies suggest a
relatively flat metallicity gradient at a Galactrocentric distance of
$R_G \ge$12\,kpc (e.g., \citealt{Korotin2014}).

\subsection{Star-forming Activities} \label{sec:SFinS208}

Before discussing the results of our deep NIR imaging with Subaru,
we discuss the star-forming activities in S208 based on 2MASS
\citep{Skrutskie2006} NIR data and {\it WISE} \citep{Wright2010} MIR
data.
Figure~\ref{fig:3col_2MASS_WISE} shows a pseudocolor image of S208 with
a field of view of $\sim$10$'\times$10$'$ and with the center at $(l, b)
= (151.29^\circ, +1.97^\circ$) in the Galactic coordinates.
The figure is produced by combining the 2MASS $K_S$-band (2.16\,$\mu$m,
blue), {\it WISE} band 1 (3.4\,$\mu$m; green), and {\it WISE} band 3
(12\,$\mu$m; red) images. 
We also show the 1.4\,GHz radio continuum from the NRAO VLA Sky Survey
(NVSS; \citealt{Condon1998}) with white contours.
{The 12\,$\mu$m emission is mainly from PAH emission}, tracing
photodissociation regions around \ion{H}{2} regions, whereas the radio
continuum traces the photoionized \ion{H}{2} region.
The distributions of the 12\,$\mu$m emission and radio continuum show
that the \ion{H}{2} region extends almost spherically with an
approximately 1\,arcmin radius. 
From the nearly perfect spherical shape centered on GSC 03719-00517,
this late O-type or early B-type star should be the main exciting source 
of the \ion{H}{2} region.

\section{OBSERVATION AND DATA REDUCTION}

\subsection{Subaru MOIRCS JHK Imaging} \label{sec:obs_MOIRCS}

Deep $JHK_S$-band images were obtained for each band with the 8.2\,m
Subaru telescope equipped with a wide-field NIR camera and spectrograph,
MOIRCS \citep{{Ichikawa2006},{Suzuki2008}}.
MOIRCS employs two 2K ``HAWAII-2'' imaging arrays, which yield a $4'
\times 7'$ field of view ($3.5' \times 4'$ for each chip) with a pixel
scale of $0''.117$ pixel$^{-1}$.
The instrument uses the MKO NIR photometric filters
\citep{{Simons2002},{Tokunaga2002}}.

The observations were performed on three nights: 2006 November 8 UT,
2007 November 23 UT, and 2008 January 14 UT.
Only on 2007 November 23 UT, the observing conditions were photometric.
On 2006 November 8 UT, it was highly humid ($\sim$45--75\,\%), while 
cirrus was sometimes seen on 2008 January 14 UT. 
Because the detector output linearity is not guaranteed for counts over
$\sim$20,000\,ADU, we obtained short-exposure images in addition to
long-exposure images for more sensitive detection.
The exposure times for the long-exposure images are 120, 15, and 30\,s
for the {\it J}, {\it H}, and $K_S$ bands, respectively, whereas the
exposure time for short-exposure images is 13\,s for all bands. 
The total integration times for the long-exposure images are 720, 420,
and 960\,s for the {\it J}, {\it H}, and $K_S$ bands, respectively,
whereas the total integration time for the short-exposure images is
52\,s for all bands.
The center of the images for S208 is set at $\alpha_{\rm 2000} = 04^{\rm
h} 19^{\rm m} 45^{\rm s}$, $\delta_{\rm 2000} = +53^\circ 05' 41''$,
which covers the whole \ion{H}{2} region described in
Section~\ref{sec:SFinS208} (see the white box in
Figure~\ref{fig:3col_2MASS_WISE} for the MOIRCS field).
For background subtraction, we also obtained images of the sky, which is
just to the north of the images for S208 by 5.5\,arcmin, to avoid the
nebulosity of S208.
We summarize the details of the observation in Tab.~\ref{tab:LOG}.

\subsection{Data Reduction and Photometry} \label{sec:Data} 

All data in each band were reduced using IRAF\footnote{IRAF is 
distributed by the National Optical Astronomy Observatories, which are
operated by the Association of Universities for Research in Astronomy,
Inc., under cooperative agreement with the National Science Foundation.}
with standard procedures, including flat fielding, bad-pixel correction,
median-sky subtraction, image shifts with dithering offsets, and image
combination.
We used sky flats made from the archived MOIRCS data in
SMOKA\footnote{SMOKA is the Subaru--Mitaka--Okayama--Kiso Archive System
operated by the Astronomy Data Center, National Astronomical Observatory
of Japan.}. We selected the data of the closest run.
In addition to the above standard procedures, distortion correction was
applied before image combination using the  
``MCSRED''\footnote{\url{http://www.naoj.org/staff/ichi/MCSRED/mcsred\_e.html}} 
reduction package for the MOIRCS imaging data. 
We constructed a pseudocolor image of S207 by combining the
long-exposure images for $J$ (1.26\,$\mu$m, blue), $H$ (1.64\,$\mu$m,
green), and $K_S$ (2.15\,$\mu$m, red) bands
(Figure~\ref{fig:3col_S208}).

$JHK_S$ photometry was performed using the IRAF apphot package for stars
in the north-east half-frame of the image for S208 (``S208 frame,''
hereafter), where a star-forming cluster was detected 
(Figure~\ref{fig:3col_S208}; see Section~\ref{sec:S208cluster}).
As photometric standards, 2MASS stars in these fields were used after
converting the 2MASS magnitudes to the MKO magnitudes using the color
transformations in \citet{Leggett2006}.

Only 2MASS stars with good 2MASS photometric quality and with colors of
$J-H \le 0.5$, $H-K_S \le 0.5$, and $J-K_S \le 1.0$ were used to avoid
the color term effect.
Because the cluster is very crowded, we used an aperture diameter of
$0''.7$ for cluster members to avoid contamination of adjacent stars.
In the IRAF apphot package, only pixel-to-pixel standard deviation in
the background sky region for each source is considered for the
magnitude errors.
In addition, we consider the flux uncertainty in the $0''.7$ aperture
that is estimated for each frame from the standard deviation of the flux
in about 3000 independent apertures in the blank area in the frame.
As a result, the limiting magnitudes (10$\sigma$) of long-exposure
images are $J=19.8$\,mag, $H=18.5$\,mag, and $K_S=18.0$\,mag.
Because the detection completeness of stars with $<$10$\sigma$ detection
is less than one, whereas that of the brighter stars is almost one (see
\citealt{Yasui2008} and \citealt{Minowa2005}),
we used only stars with all $JHK_S$ magnitudes brighter than the
limiting magnitudes (10$\sigma$) in the following.

\section{A young embedded cluster in S208} \label{sec:results} 

\subsection{Identification of a Young Cluster in S208}  
\label{sec:S208cluster}

In the pseudocolor image of the observed field with MOIRCS
(Figure~\ref{fig:3col_S208}),
we detected a star cluster in the central region of S208 from the
enhancement of stellar density compared to that of the surrounding area,
which is reported in \citet{Bica2003}.
In addition, we found a small stellar association (hereafter, S208
association) located $\sim$1.5\,arcmin toward the southwest of the
center of S208. We identified the position of the association by eye and
show it with a yellow dashed circle.
Both the cluster and the association are located near the region where
the {\it WISE} band 3 (12\,$\mu$m) emission is very strong
(Figure~\ref{fig:3col_2MASS_WISE}); this combination is often seen in
clusters (see \citealt{Koenig2012}).
Because the cluster is only seen in the northwest part of the image, and
because an engineering detector was used for the other half of the
observation of long-exposure
images\footnote{\url{http://www.subarutelescope.org/Observing/Instruments/MOIRCS/}
\url{inst\_detector.html}},
we used stars only in the northwest half-frame (the S208 frame) in the
following discussion of this section.
First, we defined the cluster region. 
We set many circles with 50 pixel ($\sim$6$''$) radius in the S208 frame
with 1 pixel step and counted the numbers of stars included in all
circles (2$\pm$2 stars on average).
From these circles, we picked a circle that contains the maximum number
of stars (approximately 30) to define the center of the cluster with an
accuracy of $\sim$5$''$: $\alpha_{\rm 2000} = 04^{\rm h} 19^{\rm m}
32.7^{\rm s}$, $\delta_{\rm 2000} = +52^\circ 58' 34.6''$.
Figure~\ref{fig:profile_S208} shows the radial variation of the
projected stellar density using stars with $K_S$ magnitudes of
$\le$18.0\,mag, corresponding to the 10$\sigma$ detection limit.
The horizontal solid line indicates the density of the region located by
more than 500\,pixel from the center of the S208 frame. 
We defined the cluster region with a circle having a radius of
300\,pixel ($34''$), where the stellar density is more than that of the 
entire sky frame by 3$\sigma$.
This is consistent with the cluster size estimated in \citet{Bica2003}
for the 2MASS data (both major and minor angular dimensions of $1'.2$).
The defined cluster region is shown as the yellow circle in
Figure~\ref{fig:3col_S208}. 
The cluster radius corresponds to 1.6\,pc and 0.7\,pc with distances of
 $D=9$\,kpc and $D=4$\,kpc, respectively. 
We also defined a control field in the S208 frame that is the area
located by more than 500\,pixels from the center of the cluster region.

\subsection{Color--Magnitude Diagram} \label{sec:CM}

We constructed the $J-K_S$ versus $K_S$ color--magnitude diagrams of
detected point sources in the S208 frame (Figure~\ref{fig:colmag_S208}).
The dwarf star tracks in spectral types of O9 to M6 (corresponding mass
of $\sim$0.1--20\,$M_\odot$) by \citet{Bessell1988} are shown as black
lines, whereas the isochrone models for the age of 1\,Myr are shown as
blue lines.
The isochrone models are by \citet{Lejeune2001} for the mass of $7 <
 M/M_\odot \le 40$, by \citet{Siess2000} for the mass of $3 < M/M_\odot
 \le 7$, and by \citet{{D'Antona1997},{D'Antona1998}} for the mass of 
 $0.017 \le M/M_\odot \le 3$.
Distances of 9\,kpc and 4\,kpc are assumed.  The arrow shows the
reddening vector of $A_V = 5$\,mag.
In the color--magnitude diagram, the extinction $A_V$ of each star was
estimated from the distance between its location and the isochrone
models along the reddening vector. 
For convenience, the isochrone model is approximated as a straight line,
shown in solid gray.
We then constructed the distributions of the extinctions of stars in the
cluster region (black) and in the control field (gray) in
Figure~\ref{fig:av_S208}.
The distribution for the control field is normalized to match the total
area of the cluster regions.
The resultant distribution for the control field shows a peak at $A_V
\sim 2$\,mag, whereas that for the cluster region shows a peak at the
much larger extinction of $A_V \sim 6$--8\,mag.
Because the differences between the two distributions are significant,
cluster members can be distinguished from contamination stars in the
cluster region based on the values of $A_V$, as in the case with the
Cloud 2 clusters \citep{Yasui2009}. 
We found that stars with $A_V \geq 4.0$\,mag are concentrated on the
cluster region, while stars with $A_V < 4.0$\,mag are widely distributed
over the observed field.
The following criteria are applied to identify members of the S208
cluster: (1) they are distributed in the cluster regions,
and (2) have large $A_V$ excess compared with normal field stars
(extinction of $A_V \ge 4$\,mag).
Actually, on the color-magnitude diagram, the cluster members are
located in the region clearly separated from the region for normal field
stars. In Figure~\ref{fig:colmag_S208}, the identified cluster members
are shown by red circles while all other sources in the S208 frame are
shown by black dots.
As a result, 89 sources ($N_{\rm cl}$) are identified as S208 cluster 
members.
The average $A_V$ value of the cluster members is estimated at $A_V =
10.1 \pm 3.9$\,mag.

Considering the relatively large $R_G$ of the S208 clusters ($R_G \ge
12$\,kpc), the contamination of background stars can be negligible and
most of the field objects are foreground stars (some are background
galaxies). To quantify the contamination, we compared the $A_V$
distributions of all the sources in the cluster regions and the field
objects in the control field (Figure~\ref{fig:av_S208}).
Because the number of field objects decreases significantly at $A_V \ge
4$\,mag, most cluster members can be distinguished from the field
objects as red sources with $A_V \ge 4$\,mag. The contamination by the
foreground stars is estimated at $\sim$2\,\% by counting the normalized
number of field objects in the tail of the distribution at $A_V \ge
4$\,mag and dividing it by the total number of sources in the cluster
regions.
In contrast, there must be some cluster members at $A_V < 4$\,mag that
missed our identification.
Using the normalized number of field objects with $A_V \ge 4$\,mag
($N'_{\rm fi}$), the number of stars in the cluster region with $A_V \ge
4$\,mag ($N'_{\rm cl}$), and the number of identified cluster members
($N_{\rm cl}$), the contamination is estimated to be $\sim$6\,\% from
$(N'_{\rm cl} - N'_{\rm fi}) / (N_{\rm cl} + (N'_{\rm cl} - N'_{\rm
fi}))$.

We placed the short horizontal lines on the isochrone models shown in
the same colors as the isochrone tracks, which show the positions of
0.1, 1, 3, and 10\,$M_\odot$.
Assuming the average $A_V$ of 10.1\,mag, the {\it K}-band limiting
magnitude of 18.0\,mag (10$\sigma$) for the age of 1\,Myr corresponds to
masses of 0.9\,$M_\odot$ and 0.2\,$M_\odot$ in the cases of photometric
distance $D=9$\,kpc and kinematic distance $D=4$\,kpc, respectively.
In any case, the mass detection limit is sufficiently low, down to the
substellar mass, which enables us to estimate the age using KLF
(Section~\ref{sec:Age_Distance}) and to derive the disk fraction with
the same criteria as in the solar neighborhood (Section~\ref{sec:DF}).
Because the most likely age and distance of S208 are estimated at
$\sim$0.5\,Myr and $\sim$4\,kpc in Section~\ref{sec:Age_Distance},
respectively, the mass detection limit is then $\sim$0.2\,$M_\odot$.

\subsection{Color--Color Diagram} \label{sec:CC} 

In Figure~\ref{fig:CC_S208}, we show the $J-H$ versus $H-K_S$
color--color diagram for stars in the S208 frame.
The cluster members are shown as red circles, while all other sources
are shown as black dots.
The dwarf star track in the late B to M6 spectral types in the MKO
system by \citet{Yasui2008} is the solid curve.
The classical T Tauri star (CTTS) locus, originally derived by
\citet{Meyer1997} in the CIT system, is shown as a gray line in the MKO
system \citep{Yasui2008}.
The arrow shows the reddening vector of $A_V = 5$\,mag.
Stars with circumstellar dust disks are known to show a large $H-K$
color excess (e.g., \citealt{Lada1992}).

The intrinsic $(H-K)$ colors ($(H-K)_0$) of each star were estimated by
dereddening along the reddening vector to the young star locus in the
color--color diagram (see Figure~\ref{fig:CC_S208}). For convenience,
the young star locus was approximated by the extension of the CTTS
locus, and only stars that are above the CTTS locus were used.
We constructed intrinsic $H-K$ color distributions for the S208 cluster
members and for those in the control field in Figure~\ref{fig:HK0_S208}.
The distribution for the control field is normalized to match the total
area of the cluster regions.
The distribution of the cluster members seems to have a larger number of
red stars with $(H-K)_0 > 0.2$\,mag compared to those in the control
field.  The average $(H-K)_0$ value for cluster members is estimated at
0.35\,mag, whereas that in the control field is estimated at 0.23\,mag.
The difference in the average $(H-K)_0$ between the stars in the cluster
region and in the field stars (0.12\,mag) can be attributed to thermal
emission from the circumstellar disks of the cluster members. Assuming
that disk emissions appear in the $K$ band but not in the {\it H} band,
the disk color excess of the S208 cluster members in the {\it K} band,
$\Delta K_{\rm disk}$, is equal to 0.12\,mag.

\subsection{K-band Luminosity Function (KLF)} \label{sec:KLF}

We constructed the KLF for the S208 cluster members (black line in
Figure~\ref{fig:KLFobs_S208}).
The number counts of the KLF generally increase in the fainter magnitude
bins.
However, the number counts become flat in the $K =15.5$--16.5\,mag bin
and they then decrease in the $K=17.5$\,mag bin, which generally
corresponds to the peak of IMF.
Considering the 10$\sigma$ detection magnitude of $K=18.0$\,mag for the
S208 frame, the completeness should be $\sim$1 in all the magnitude bins
(Section~\ref{sec:Data}; see also \citealt{Yasui2008} and
\citealt{Minowa2005}).
Therefore, even considering the detection completeness, the S208 KLF
would have a peak at the $K=15.5$--16.5\,mag bin.

It should be noted that the large $A_V$ dispersion of the S208 cluster,
$A_V \sim 4$--25\,mag, can make it difficult to detect faint cluster
members.
This may be the cause for the decrease of the KLF in fainter magnitude
 bins.
For comparison, we also constructed KLFs for stars with limited $A_V$
values in Figure~\ref{fig:KLFobs_S208}, $A_V = 5$--15\,mag and $A_V =
6.2$--13.0\,mag (from the $A_V$ distribution of the cluster members,
10.1$\pm$3.9\,mag), which are shown by the solid and dashed gray lines,
respectively.
For clarity, the KLFs are shifted vertically by $+$0.1\,mag and
$+$0.2\,mag for the KLFs of stars with $A_V = 5$--15\,mag and $A_V =
6.2$--13.0\,mag, respectively.
As a result, the discrepancy between all three KLFs are found to be
within the uncertainty range, suggesting that the selection of stars
with different limited $A_V$ values causes a negligible influence on the
obtained KLF.
Therefore, we used the original KLF (KLF from all S208 cluster members) 
in the following discussion.

\section{Discussion} \label{sec:Discussion} 

\subsection{Age and Distance of S208} \label{sec:Age_Distance}

The KLFs of different ages are known to have different peak magnitudes
and slopes, fainter peak magnitudes and less steep slopes with
increasing age \citep{Muench2000}.
By comparing observed and model KLFs, the age of the young clusters can
be roughly estimated with an uncertainty of $\pm$1\,Myr
(\citealt{{Yasui2006},{Yasui2008}}).
We constructed model KLFs in the same way as in our previous work (see
Section~4 in \citealt{Yasui2006}) with the assumed distance of the
cluster that underlies typical IMFs and mass--luminosity relations.
We used the Trapezium IMF \citep{Muench2002}, which is considered to be
the most reliable IMF for young clusters (e.g., \citealt{LadaLada2003}),
as discussed in \citet{Yasui2008}.
We used isochrone models in Section~\ref{sec:CM}: models by
\citet{Lejeune2001} for the high-mass range ($7 < M/M_\odot \le 40$), by
\citet{Siess2000} for the intermediate-mass range ($3 < M/M_\odot \le
7$), and by \citet{{D'Antona1997},{D'Antona1998}} for the low-mass range
($0.017 \le M/M_\odot \le 3$).
We constructed model KLFs with ages of 0.1, 0.5, 1, 2, and 3\,Myr
assuming the kinematic distance ($D=9$\,kpc;
Figure~\ref{fig:KLFfit_S208}, left) and the photometric distance
($D=4$\,kpc; Figure~\ref{fig:KLFfit_S208}, right).
We also considered the $A_V$ and $\Delta K_{\rm excess}$ estimated in
Sections~\ref{sec:CM} and \ref{sec:CC}.

We compared the observed KLF (black lines in
Figure~\ref{fig:KLFfit_S208}) and model KLFs, constructed in the same
way as in \citeauthor{Yasui2006} (\citeyear{Yasui2006},
\citeyear{Yasui2008}; colored lines in Figure~\ref{fig:KLFfit_S208}),
for the two distances.
The model KLFs were normalized so that the total number of stars in the
magnitude range $K=12$--18\,mag is the same as that for the observed
KLF.
In the case of $D=9$\,kpc (Figure~\ref{fig:KLFfit_S208}, left), the peak
magnitude of the observed KLF ($K=15.5$--16.5\,mag) is found to be much
brighter than those of the model KLFs for any ages of more than 0.1\,Myr
($K \ge 18.5$\,mag).
The peak magnitude of the observed KLF corresponds to $\sim$3\,$M_\odot$
for the model KLF with the age of 0.1\,Myr. The corresponding mass
becomes even larger for older ages, which is obviously implausible for
the peak mass of the underlying IMF.
Therefore, the quoted photometric distance, $D=9$\,kpc, does not seem to 
be correct.
On the other hand, in the case of $D=4$\,kpc
(Figure~\ref{fig:KLFfit_S208}, right), the model KLFs with ages of
0.1--0.5\,Myr appear to fit the observed KLF for both the peak and slope
of the KLF.
However, it is difficult to estimate the age of the cluster with an
accuracy of 0.1\,Myr because isochrone models for ages of less than
1\,Myr are thought to be uncertain (e.g., \citealt{Baraffe2002}).
Therefore, the age of $\sim$0.5\,Myr and the distance of $D=4$\,kpc are
most likely parameters for the S208 cluster.
The suggested distance of the S208 cluster is consistent with the
kinematic distance and with S208 being located in the outer arm, as
described in Section~\ref{sec:properties}.
This is also the case for S207 (see \citealt{Yasui2016}), except that
S207 is slightly older ($\sim$2--3\,Myr). Therefore, both \ion{H}{2}
regions appear to make a combination in the outer arm. 
We will discuss the star formation for those two \ion{H}{2} regions in 
the outer arm in Section~\ref{sec:SFinOA}.


\subsection{Implication for IMF} \label{sec:IMF}

CO emission was detected in S208 by \citet{Blitz1982} and
\citet{Wouterloot1989} as described in Section~\ref{sec:properties}.
Although the CO emission by \citet{Wouterloot1989} does not cover the
cluster region, that by \citet{Blitz1982} totally covers the
region\footnote{The central position of the CO emission in
\citet{Wouterloot1989} is more than 100$''$ apart from the IRAS
positions (plus symbol in Figure~\ref{fig:3col_2MASS_WISE})
with the beam size of their observation (21$''$), while that by
\citet{Blitz1982} is about 30$''$ off the center of the S208 cluster
region with the beam size of $\sim$2$'$ in their observation.}. 
Based on the physical parameters estimated by \citet{Blitz1982}, $T_A^*
= 13$\,K and $\Delta V = 9.8$\,km\,s$^{-1}$,
the H$_2$ column density\footnote{The telescopes at the Bell Telescope
Laboratories (BTL) and the Millimeter Wave Observatory (MWO) were used
for the S208 observation in \citet{Blitz1982}.
The column density is estimated considering a main beam efficiency of
89\,\% for BTL \citep{Bally1987} and $\sim$80\,\% for MWO
\citep{Magnani1985}.} is estimated to be
$\sim$4--5$\times$10$^{22}$\,cm$^{-2}$, which corresponds to $A_V =
50$\,mag.
Considering that the molecular gas disperses on a timescale of 
$\sim$3\,Myr \citep{Hartmann2001}, the age of the S208 cluster is at
least $\lesssim$3\,Myr.
The high column density suggests a quite young age. 
However, for a more precise distribution of the CO molecular cloud
relative to the cluster, CO mapping with high sensitivity will be
necessary in the future.

In Section~\ref{sec:CC}, the average extinction of stars in the S208
cluster is estimated at $\langle A_V \rangle \sim 10$\,mag with a peak
of $A_V = 6$--8\,mag.
Considering $A_V \sim 3$\,mag for the foreground extinction in the
direction of S208, which was estimated from a three-dimensional 
extinction map of the Galactic plane \citep{Sale2014}, the average
intrinsic extinction of the S208 cluster is estimated at $\langle A_V 
\rangle \sim 7$\,mag.
Generally, the average extinction is higher for younger clusters: 
$\langle A_V \rangle \sim 11$\,mag for NGC 2024 ($\sim$0.5\,Myr old) by
\citeauthor{Meyer2008} (\citeyear{Meyer2008}; see also
\citealt{Haisch2000AJ}), 
$\langle A_V \rangle = 9$\,mag for Orion Trapezium ($\sim$1\,Myr old) by
\citet{Muench2002},
$\langle A_V \rangle = 3.5$\,mag for IC 348 (2.5\,Myr old) by
\citet{Herbst2008}, and $\langle A_V \rangle = 0.5$\,mag for NGC 2264
(3\,Myr old) by \citet{Rebull2002}.
The S208 cluster shows the high extinction that is similar to NGC 2024
and Trapezium, suggesting that the age of the S208 cluster is quite
young, $\sim$0.5--1\,Myr.
Because extinctions can be estimated lower considering the low
metallicity of S208 (see e.g., \citealt{Garn2010}), the actual gas mass
may be larger, implying that the S208 cluster is even younger.

In the KLF fitting (Section~\ref{sec:Age_Distance}), a typical IMF is
assumed for estimating the age of the S208 cluster.  We found that the
estimated age ($\sim$0.5\,Myr) is consistent with the above roughly
estimated age based on independent information.
This suggests that the IMF of the S208 cluster, which is located in very
low-metallicity environments, can be approximated by the typical IMFs of
the solar neighborhood ($\sim$0\,dex) for the mass range of
$\ge$0.2\,$M_\odot$.
Because the KLF slope on the bright side is very sensitive to the
higher-mass slope of the IMF \citep{{Yasui2006},{Yasui2008ASPC}}, 
the good fit of the observed KLF down to the peak at $m_K \sim 16$\,mag
suggests that the higher-mass slope of the IMF is not significantly
different from typical IMFs.
The characteristic mass ($M_c$), which is the mass of the IMF peak
corresponding to the peak of the KLF, is estimated at
$\sim$0.6\,$M_\odot$ for the age of 0.5\,Myr.
This is identical to the characteristic mass of the IMF in the solar
neighborhoood, $M_c \sim 0.3$\,$M_\odot$ ($\log M_c / M_\odot \sim -0.5 
\pm 0.5$; \citealt{Elmegreen2008}).
Similarly, the universal IMF has also been suggested in our previous
 studies of other young clusters in low-metallicity environments, Cloud
 2-N, -S clusters \citep{{Yasui2006},{Yasui2008}}, and S207 
 \citep{Yasui2016}.


\subsection{Disk Fraction} \label{sec:DF}  

The ratio of stars with protoplanetary disks in young clusters, the disk
fraction, is one of the most fundamental parameters characterizing disk
formation and evolution, and ultimately planet formation
\citep{{Haisch2001ApJL},{LadaLada2003}}.
On the {\it JHK} color--color diagram, stars without circumstellar disks
are seen as main-sequence stars reddened with extinction, whereas stars
with circumstellar disks are seen in ``the disk-excess region,'' which
is the orange highlighted region to the right of the dot--dashed line in
Figure~\ref{fig:CC_S208} because of thermal emissions from a hot dust
disk with a temperature of $\sim$1500\,K.
The dot--dashed line intersecting the dwarf star curve at maximum
$H-K_S$ values (M6 point on the curve) and is parallel to the reddening
vector is the border between stars with and without circumstellar disks
(see details in \citealt{Yasui2009}).
The disk fraction for identified cluster members with more than
10$\sigma$ detection for all $JHK_S$ bands is estimated at 27$\pm$6\,\%
(24/89) from Figure~\ref{fig:CC_S208}.

In Figure~\ref{fig:HK0_disk}, we show the fraction of stars ($f_{\rm
stars}$) per each intrinsic $(H-K)$ color bin $(H-K)_0$ for the S208
cluster (red), which is shown as the black line in
Figure~\ref{fig:HK0_S208}, and those for other young clusters in
low-metallicity environments, S207 (black solid line), Cloud 2-N (black
dashed line), and Cloud 2-S (black dotted line) with estimated disk
fractions of 4\,\%, 9\,\%, and 27\,\%, respectively \citep{Yasui2009}.
The vertical dashed line shows the borderline for estimating the disk
fraction in the MKO system.\footnote{Although $(H-K)_0 = 0.43$\,mag was
shown for the borderline in the MKO system in \citet{Yasui2009}, we
found that the correct borderline is $(H-K)_0 = 0.52$\,mag.}
The distribution becomes bluer and sharper with lower disk fractions for
nearby young clusters (see the bottom panel of Figure~7 in
\citealt{Yasui2009}), which is also the case for clusters in
low-metallicity environments \citep{Yasui2009}.
The peak $(H - K)_0$ of the S208 cluster is relatively red, $(H - K)_0
\sim 0.3$\,mag, while the distribution is relatively broad with the
maximum a $(H - K)_0$ of $\sim$2\,mag.
The distribution of the S208 cluster resembles that of the Cloud 2-S
cluster with a disk fraction of $\sim$25\,\% rather than those of the
S207 cluster and the Cloud 2-N cluster with a disk fraction of
$<$10\,\%.
The distribution of the S208 cluster is consistent with the estimated
disk fraction, 27$\pm$6\,\%.

NIR disk fractions are known to have high values ($\sim$60\,\%) for very
young clusters but decrease with increasing age, and then reach 
$\sim$5--10\,\% on a timescale of $\sim$10\,Myr (\citealt{Lada1999};
\citealt{Hillenbrand2005}; \citealt{Yasui2010}; see the red line in the
left panel of Fig~5 in \citealt{Yasui2014}).
Although NIR disk fractions are generally slightly lower than MIR disk
 fractions, which are based on ground $L$-band observations and space
 MIR observations, the characteristics are quite similar. 
As suggested in \citet{Yasui2010}, the derived disk fraction for the
S208 cluster (27$\pm$6\,\%) is lower than that for clusters in the solar
neighborhood with ages similar to the S208 cluster's age ($\sim$60\,\%
for $<$1\,Myr).
The lower disk fraction in low-metallicity environments suggests that
the disk lifetime in low-metallicity environments is quite short, as
discussed in \citet{Yasui2009}.

It should be noted that only the S208 cluster has very red stars with
 $(H - K)_0 \ge 1.5$\,mag in Figure~\ref{fig:HK0_disk}.
This may be because the cluster is the youngest among those in
Figure~\ref{fig:HK0_disk}: the estimated age of the S208 cluster is
$\sim$0.5\,Myr (Section~\ref{sec:Age_Distance}), while those of the
Cloud 2-N, Cloud 2-S, and S207 clusters are 0.5--1\,Myr, 0.5--1\,Myr,
and 2--3\,Myr, respectively (see \citealt{Yasui2016} and references
therein).
Very young YSOs that are surrounded by thick circumstellar envelopes of
falling material, called Class I sources, are known to show very red NIR
colors \citep{Kenyon1995}.
Although the classification of sources between Class I and Class II
(YSOs surrounded by optically thick circumstellar disks) cannot be
perfectly distinguished using the {\it JHK} color-color diagram,
many of the Class I sources in NGC 2024 are located in the disk-excess
region and have large $J-K$ colors (equal to the sum of $J-H$ (y-axis)
and $H-K$ (x-axis) colors) of larger than $\sim$3 (see left panel of
Figure~8 in \citealt{Haisch2000AJ}). The same trend is seen in the
Taurus star-forming region.\footnote{We checked the colors of the Class
I sources in the Taurus with NIR photometry data by \citet{Kenyon1995}
and information of classifications by \citet{Kenyon1998}.}
The S208 cluster has four objects with such colors. 
Our results suggest that YSOs in low-metallicity environments are
initially surrounded by thick circumstellar envelopes, as is the case
for the solar neighborhood, but the circumstellar material as well as
the disk disperse very quickly.


\section{Star formation in the outer arm} \label{sec:SFinOA}

\begin{figure}[h]
\begin{center}
\includegraphics[width=8.cm]{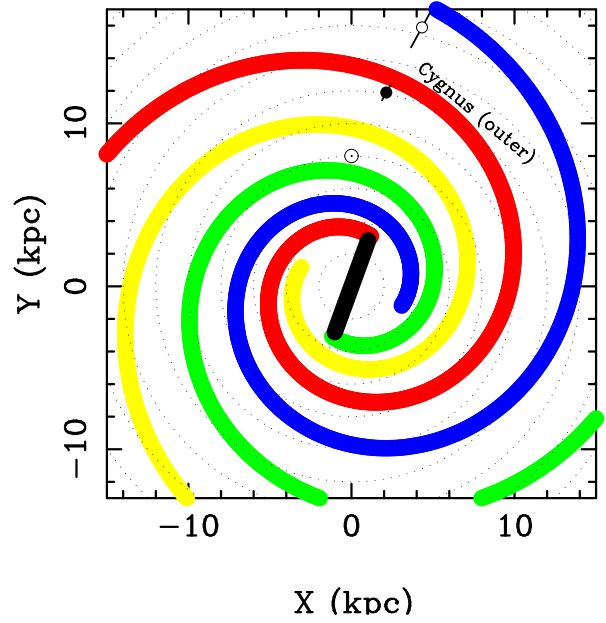}
\caption{Top view of the Milky Way Galaxy, with S208 and spiral arms. 
S208 is shown with an open circle in the case of photometric distance
 and a filled circle in the case of kinematic distance.
The photometric distance is based on all the derivations in
Section~\ref{sec:properties}, $D \sim 9$\,kpc (7.6--10\,kpc), while
the kinematic distance is based on most recent derivation by
 \citet{Foster2015}, $D=4.4$\,kpc with an uncertainty of 0.55\,kpc,
 covering all1 the previous derivations in Section~\ref{sec:properties}.
The spiral arms from \citet{Vallee2005}, which are shown with different
colors: red, yellow, green, and blue
for Norma--Cygnus (outer) arm, Perseus arm, Sagitarius--Carina arm, and
Scutum--Crux arm, respectively.
The Sun is shown by a circled dot assuming the Galactrocentric distance
 of 8.0\,kpc.
Dots show concentric circles around the Galactic center at a Galactic
 radius $r=2$, 4, 6, ..., 22\,kpc.}  \label{fig:Gal} 
\end{center}
\end{figure}

Figure~\ref{fig:3col_WISE} (left) shows a {\it WISE} MIR pseudocolor
image around S207 and S208 with a field of view of $30\,{\rm arcmin}
\times 30$\,arcmin centered at $(l, b) = (151.24^\circ, +2.05^\circ)$ in
Galactic coordinates.
S207 is located in the upper right in the figure, while S208 is located
in the lower left, $\sim$10\,arcmin southward of S207. 
Since the identical distances of S207 and S208, $\sim$4\,kpc, are
supported both from the kinematic distance and from the KLF fitting
(Section~5.1 in \citealt{Yasui2016} for S207, and
Section~\ref{sec:Age_Distance} for S208), the actual separation is
suggested to be very small, 12\,pc.
Based on the distance, their Galactrocentric distances are estimated at
$R_G \simeq 12$\,kpc, suggesting that both clusters are located in the
outer arm (Figure~\ref{fig:Gal}).
The fact that the kinematic distance is consistent with the photometric
(KLF) distance may support the suggestion that sources in the outer arm
are in nearly circular Galactic orbits (e.g., \citealt{Hachisuka2015}).  

\begin{figure*}[!h]
\begin{center}
\includegraphics[width=12cm]{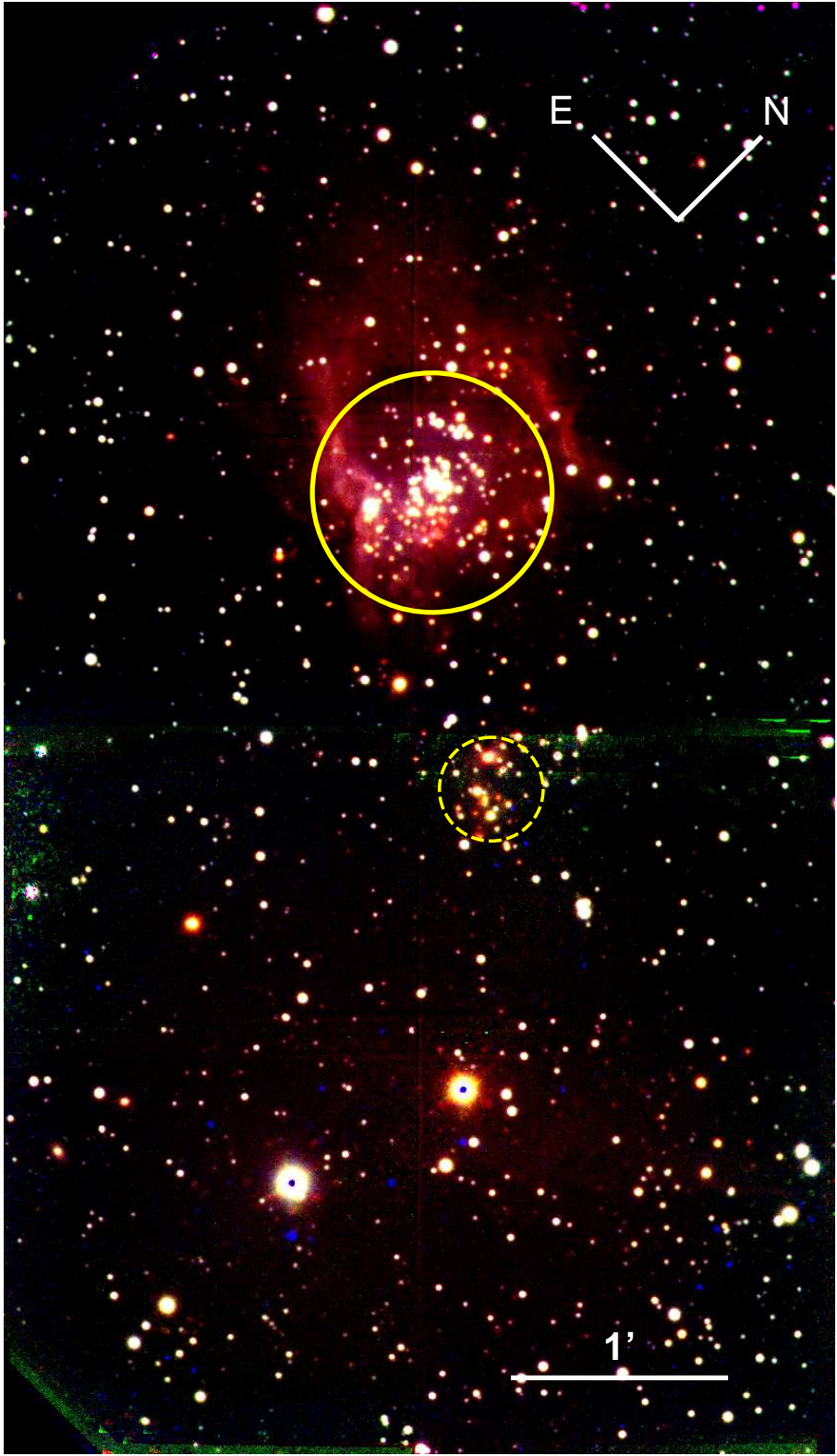}
\caption{$JHK_S$ pseudocolor image of S208.  The color image is produced
by combining the $J$- (1.26\,$\mu$m), $H$- (1.64\,$\mu$m), and
$K_S$-band (2.15\,$\mu$m) images obtained with MOIRCS at the Subaru
telescope on 2008 January for $J$ band and on 2007 November for $H$ and
$K_S$ bands with the center of $\alpha_{\rm 2000} = 04^{\rm h} 19^{\rm
m} 45^{\rm s}$, $\delta_{\rm 2000} = +53^\circ 05' 41''$ in Equatorial
coordinates.
The field of view of the image is $\sim$7$'\times$4$'$, which is shown
 with a white box in Fig.~\ref{fig:3col_2MASS_WISE}. 
The yellow circle ($r=35''$) shows the location of the cluster with the
central coordinate of $\alpha_{\rm 2000} = 04^{\rm h} 19^{\rm m}
32.7^{\rm s}$, $\delta_{\rm 2000} = +52^\circ 58' 34.6''$.
The yellow dashed circle ($r=15''$) shows the location of the small
stellar association, which is identified by eye, with the central
coordinate of $\alpha_{\rm 2000} = 04^{\rm h} 19^{\rm m} 24.6^{\rm s}$,
$\delta_{\rm 2000} = +52^\circ 57' 50.8''$.}  \label{fig:3col_S208}
\end{center}
\end{figure*}

\begin{figure}[!h]
\begin{center}
\includegraphics[width=8.cm]{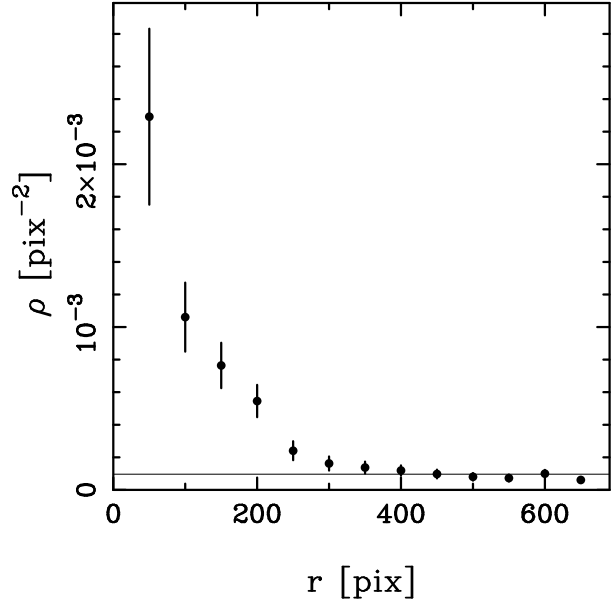}
\caption{Radial variation of the projected stellar density of stars
(filled circles) in the S208 cluster region with the center of
{$\alpha_{\rm 2000} = 04^{\rm h} 19^{\rm m} 32.7^{\rm s}$, $\delta_{\rm
2000} = +52^\circ 58' 34.6''$.}
50\,pixels correspond to $\sim$6$''$. 
The error bars represent Poisson errors.  The horizontal solid line
denotes the star density in the control field.}
\label{fig:profile_S208}
\end{center}
\end{figure}

\begin{figure}[!h]
\begin{center}
\includegraphics[width=8cm]{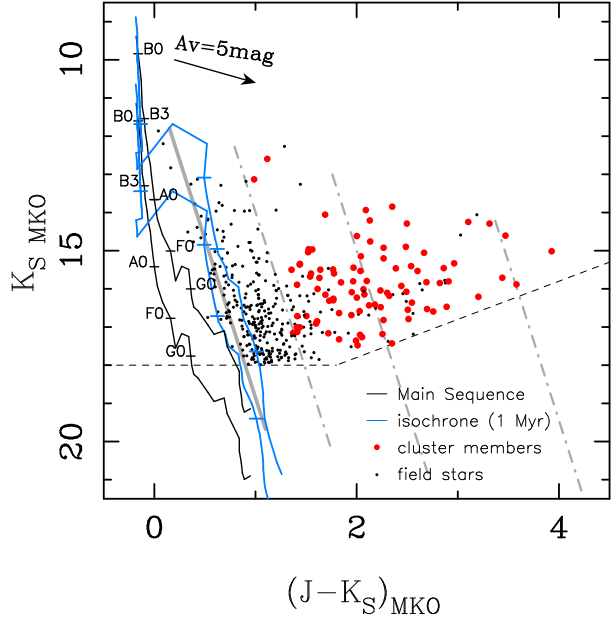}  
\caption{$(J - K_S)$ vs. $K_S$ color--magnitude diagram of S208.
Identified cluster members in the cluster region (yellow circle in
Fig.~\ref{fig:3col_S208}) are shown with red filled circles, while other
sources are shown with black dots.
The arrow shows the reddening vector of $A_V = 5$\,mag. 
The dashed lines mark the limiting magnitudes (10$\sigma$).  
The black lines show the dwarf tracks by \citet{Bessell1988} in the
spectral type of O9 to M6 (corresponding mass of
$\sim$0.1--20\,$M_\odot$).
The blue line denotes the isochrone models for the age of 1\,Myr old by
\citeauthor{D'Antona1997} (\citeyear{D'Antona1997},
\citeyear{D'Antona1998}; $0.017 \le M/M_\odot \le 3$),
\citeauthor{Siess2000} (\citeyear{Siess2000}; $3 < M/M_\odot \le 7$),
and \citeauthor{Lejeune2001} (\citeyear{Lejeune2001}; $7 < M/M_\odot \le
40$).
The distances of 9\,kpc and 4\,kpc are assumed.
The short horizontal lines are placed on the isochrone models and are
shown with the same colors as the isochrone tracks, which show the
positions of 0.1, 1, 3, and 10\,$M_\odot$.
For convenience, the isochrone models are approximated as straight
lines, shown as the solid gray, for estimating the $A_V$ value for each
star.
The dotted--dashed gray lines show the approximated isochrone models
with extinctions of $A_V = 4$, 10, and 20\,mag.}
\label{fig:colmag_S208}
\end{center}
\end{figure}

\begin{figure}[!h]
\begin{center}
\includegraphics[width=8cm]{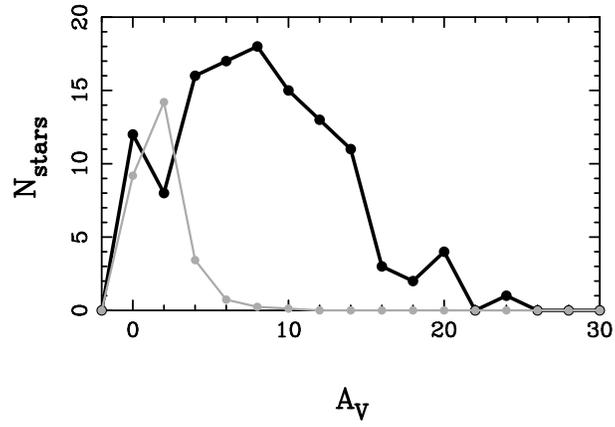}
\caption{$A_V$ distributions of stars in the S208 cluster region (black
line) and stars in the control field (gray line).  The distribution for
the control field is normalized to match the total area of the cluster 
region.}
\label{fig:av_S208}
\end{center}
\end{figure}

\begin{figure}[h]
\begin{center}
\includegraphics[width=8cm]{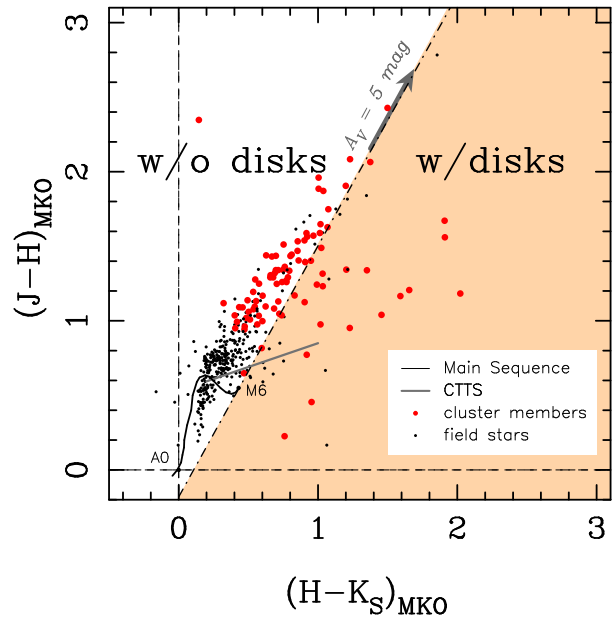}
\caption{$(H-K_S)$ vs. $(J - H)$ color--color diagram of S208.  
Identified cluster members are shown as red filled circles, while all
other stars in the S208 frame are shown as black dots.
The solid curve in the lower left portion is the locus of the points
corresponding to unreddened main-sequence stars.  The dot--dashed line,
which intersects the main-sequence curve at the maximum {\it H}$-$$K_S$ 
values (M6 point on the curve) and is parallel to the reddening vector,
is the border between stars with and without circumstellar disks.
The classical T Tauri star (CTTS) locus is shown with the gray line.} 
\label{fig:CC_S208}
\end{center}
\end{figure}

\begin{figure}[!h]
\begin{center}
\includegraphics[width=8cm]{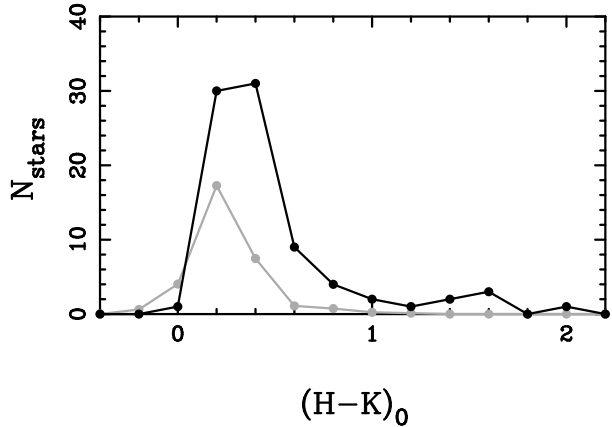}
\caption{$(H-K)_0$ distributions for the S208 cluster members (black
line) and stars in the control field (gray line).
The distribution for the control field is normalized to match the total
area of the cluster region. }
\label{fig:HK0_S208}
\end{center}
\end{figure}

\begin{figure}[!h]
\begin{center}
\includegraphics[width=8cm]{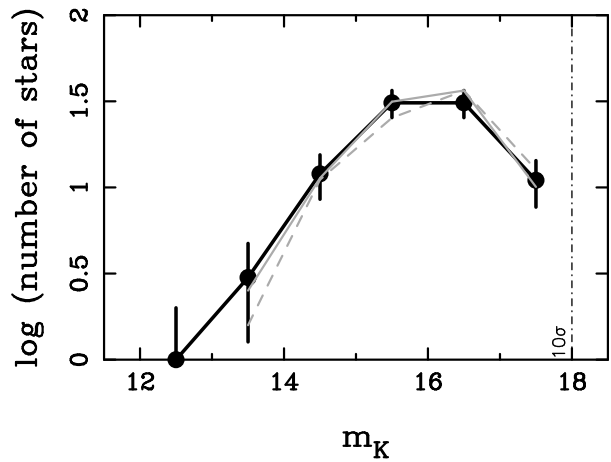}
\caption{Raw KLFs for the S208 cluster members.
The KLF for all the cluster members is shown by the black line with
 black dots ($A_V \ge 4$\,mag).  Error bars are the uncertainties from
 Poisson statistics.
The KLFs for limited $A_V$ samples are shown by gray lines, the gray
solid line for the cluster members with $A_V$ of 5--15\,mag and the gray
dashed line for the members with $A_V$ of 6.2--13.0\,mag.
The gray solid line and gray dashed line are vertically shifted by $+$0.1
and $+$0.2, respectively.
The vertical dotted--dashed line shows the limiting magnitudes of the
 10\,$\sigma$ detection (18.0\,mag).} 
\label{fig:KLFobs_S208} 
\end{center}
\end{figure}

\begin{figure*}[t]
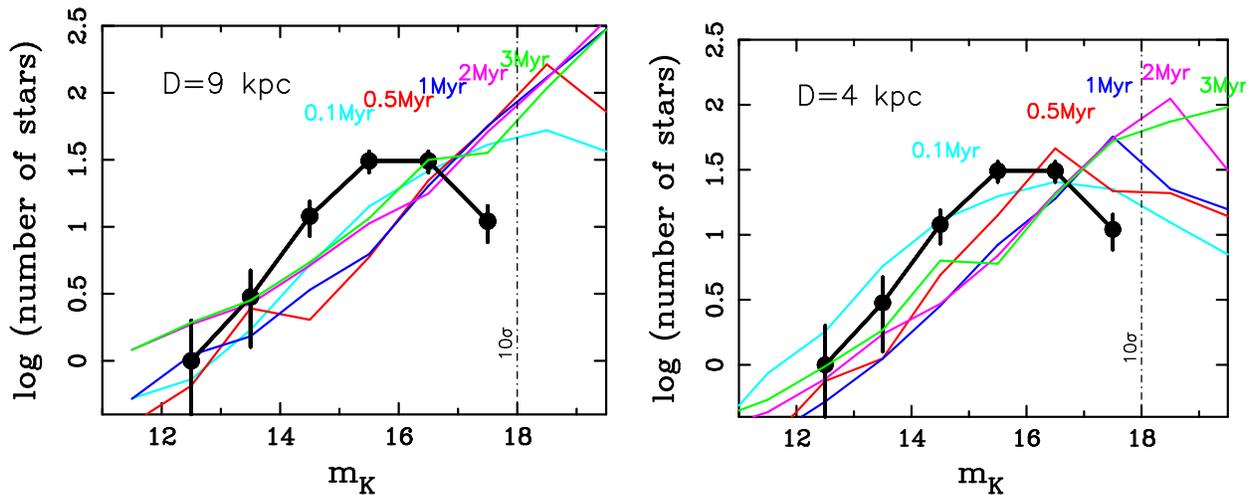

\begin{center}
\includegraphics[width=8cm]{KLF_S208_2015Oct_05_LeSiDA_D9_10sig.eps}
\hspace{1em}
\includegraphics[width=8cm]{KLF_S208_2015Oct_05_LeSiDA_D4_10sig.eps}
\caption{Comparison of the S208 KLFs (black lines) with model KLFs of
various ages (colored lines).
Error bars are the uncertainties from Poisson statistics. 
Two cases for the distance are assumed: photometric distance $D=9$\,kpc
(left) and kinematic distance $D=4$\,kpc (right). 
The aqua, red, blue, magenta, and green lines represent model KLFs of
0.1, 0.5, 1, 2, and 3\,Myr, respectively. 
The vertical dot--dashed lines show the limiting magnitudes of the
10$\sigma$ detection (18.0\,mag).} \label{fig:KLFfit_S208}
\end{center}
\end{figure*}

\begin{figure}[!h]
\begin{center}
\includegraphics[width=8cm]{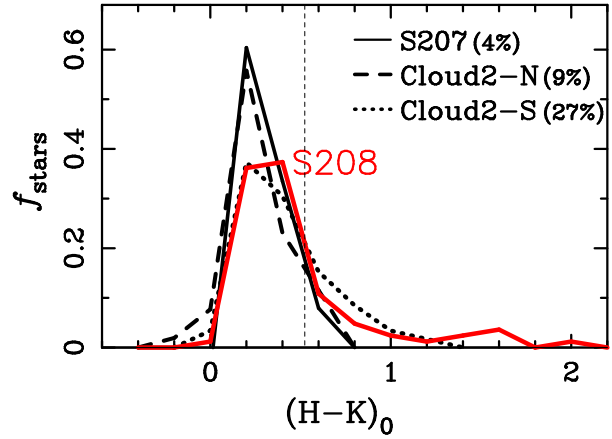}
\caption{Comparison of intrinsic $H-K$ color distributions. The
 fractions of stars ($f_{\rm stars}$) per each intrinsic color bin
 $(H-K)_0$ for clusters in low-metallicity environments, S208, S207,
 Cloud 2-N, and Cloud 2-S are plotted. The vertical dashed line shows
 the borderline for estimating the disk fraction in the MKO system.}
 \label{fig:HK0_disk}
\end{center}
\end{figure}

\begin{figure*}[h]
\begin{center}
\includegraphics[width=8.5cm]{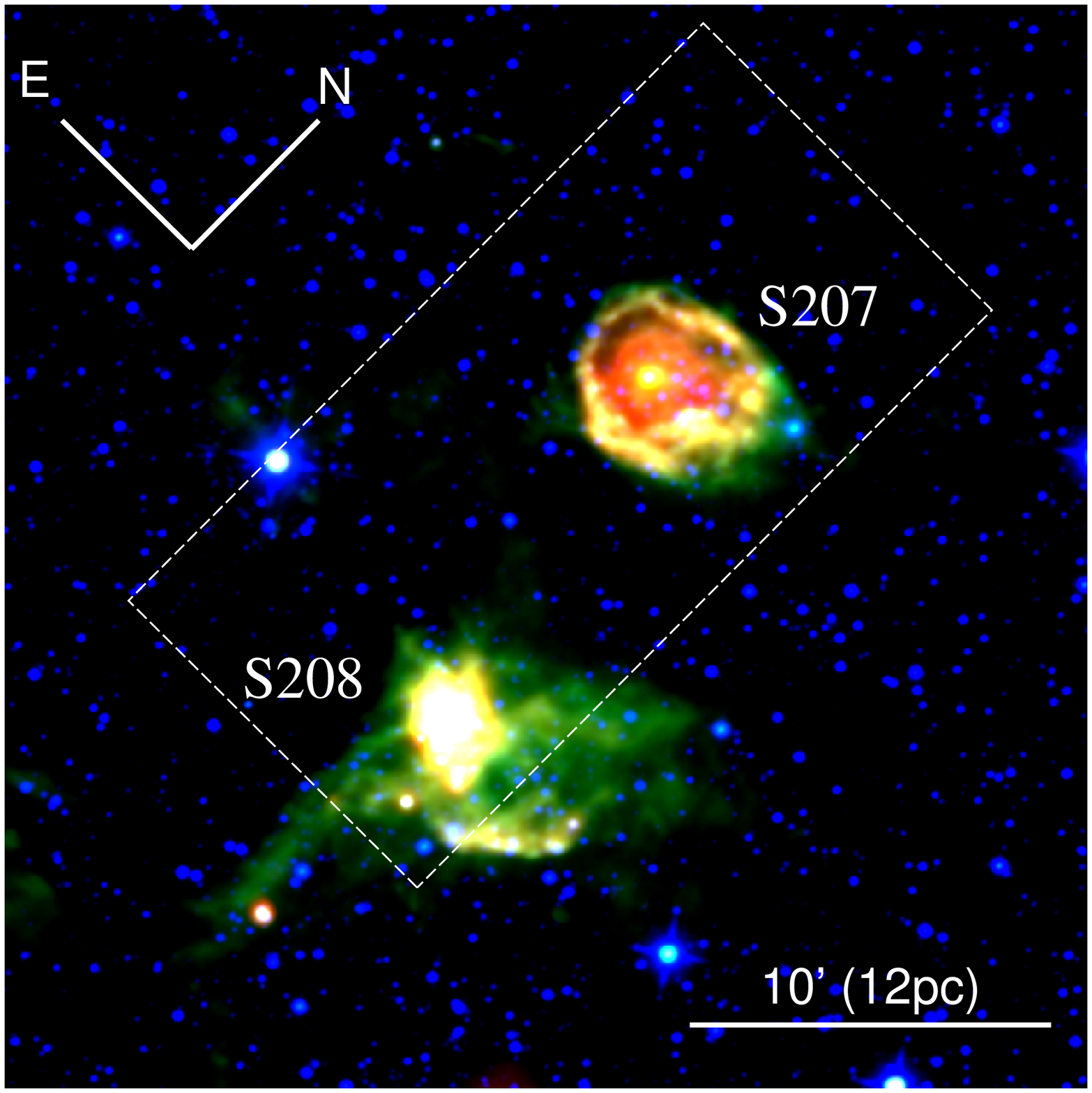}
\includegraphics[width=8.5cm]{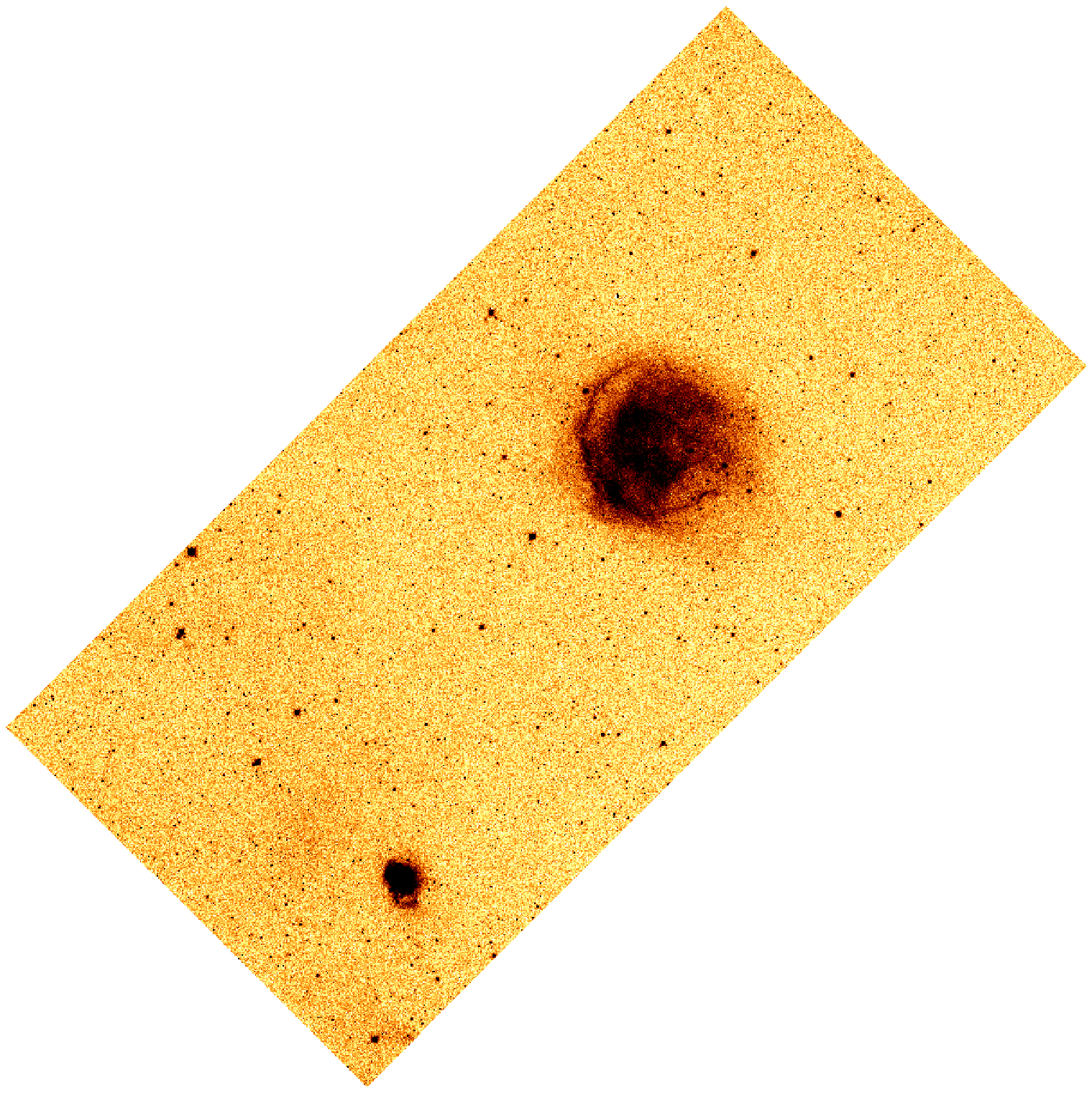}
\caption{{\it Left}: {\it WISE} MIR pseudocolor image of S207 and S208
with a field of view (30$\,{\rm arcmin} \times 30$\,arcmin) centered at
$(l, b) = (151.24^\circ, +2.05^\circ$) in Galactic coordinates.
Galactic longitude is along the x-axis, while Galactic latitude is along
 the y-axis. 
The 10\,arcmin corresponds to 12\,pc for the distance of 4\,kpc.
The image is produced by combining {\it WISE} band 1 (3.4\,$\mu$m; 
blue), {\it WISE} band 3 (12\,$\mu$m; blue), and {\it WISE} band 4
(22\,$\mu$m; red).
{\it Right}: IPHAS H$\alpha$ image of S207 and S208 with the same scale
 as the left panel. The field of view is
 $\sim$11.3\,arcmin$\times$22.6\,arcmin, which is shown with a white
 dashed box in the left panel.}  \label{fig:3col_WISE}
\end{center}
\end{figure*}

Because the estimated metallicities of both clusters are also almost
identical, $-$0.8\,dex, the star-formation activities in the two
clusters are suggested to originate in a similar environment, except
that they are at different evolutionary stages: S207 is suggested to be
at the end of the embedded cluster phase ($\sim$2--3\,Myr), while S208
is suggested to be in a very young phase ($\sim$0.5\,Myr) from the KLF
fitting.
This is also suggested from the NIR pseudocolor images, the bluer colors
of the S207 cluster (Figure~1, bottom in \citealt{Yasui2016}), and the
redder colors of the S208 cluster (see Figure~\ref{fig:3col_S208}).
The differences in the following two star-forming properties of the star
 formation support the difference in the cluster ages.
{(i) Cluster size.} The cluster size is suggested to be larger with
increasing ages due to proper motions of individual stars
\citep{Stahler2005}.
This is the case for the S207 and S208 clusters: the sizes are about
1.3\,pc and 0.7\,pc in radius, respectively.
These radii are consistent with those in the cluster samples of
\citet{Pfalzner2009}, in which many of them are nearby clusters (see
Figure~2b in \citealt{Pfalzner2009}).
(ii) Size of an \ion{H}{2} region.  Size of an \ion{H}{2} region is
known to become larger with an increasing age \citep{Dyson1980}.
This is the case for S207 and S208: the radius of 3\,pc for S207 and
 that of 1.2\,pc for S208 from the 1.4\,GHz radio continuum.
An H$\alpha$ image (Figure~\ref{fig:3col_WISE}, right) from the IPHAS
survey \citep{Drew2005} around S207 and S208 also shows the clear
difference in size of the two \ion{H}{2} regions.
\citet{Dyson1980} inferred the age of the \ion{H}{2} region with the
assumption of a uniform pressure within the ionized region.
According to this model, the age is estimated to be 0.2\,Myr for S207
and 0.1\,Myr for S208 assuming an original ambient density of $n=10^3$
cm$^{-3}$ and a flux of ionizing photons for the \ion{H}{2} regions
($N_{\rm Ly}$) of $1.25 \times 10^{48}$\,s$^{-1}$ emitted from a B0V
star (\citealt{Vacca1996}; \citealt{Schaerer1997}).
Although this age estimate is crude because the assumption of evolution
in a strictly uniform medium is unrealistic \citep{Deharveng2006}, the
trends in the size of the \ion{H}{2} region for S207 and S208 should
reflect the age sequence.

\begin{figure}[h]
\begin{center}
\includegraphics[width=15.5cm]{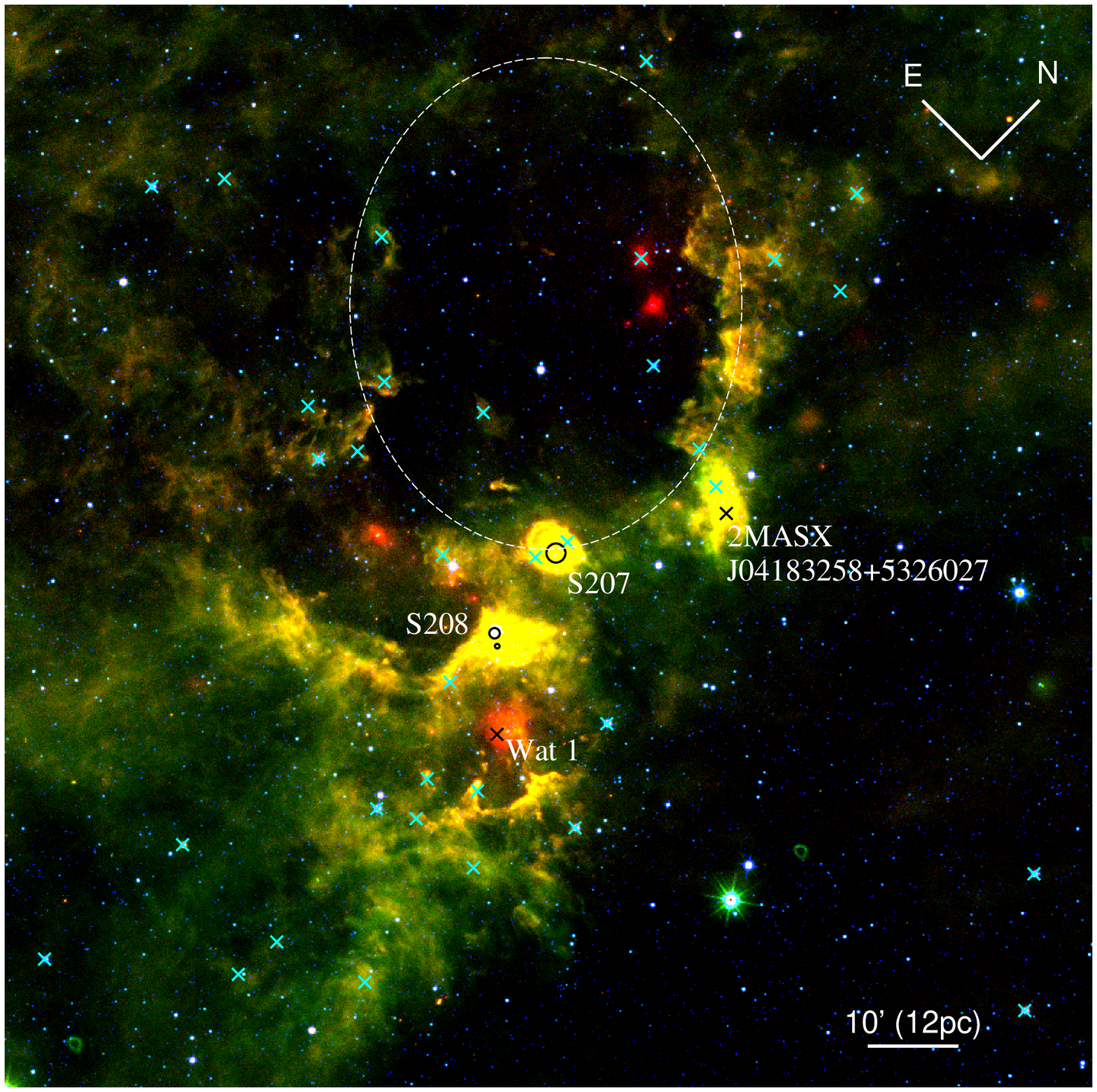}
\caption{MIR pseudocolor image of S207 and S208 with a wide field of
 view (2$\,{\rm deg} \times 2$\,deg) centered at $(l, b) =
 (151.19^\circ, +2.13^\circ$) in Galactic coordinates.
Galactic longitude is along the x-axis, while Galactic latitude is along
the y-axis. 
 The 10\,arcmin corresponds to 12\,pc for the distance of 4\,kpc.  The
 image is produced by combining the {\it WISE} band 1 (3.4\,$\mu$m;
 blue), {\it WISE} band 3 (12\,$\mu$m; blue), and {\it WISE} band 4
 (22\,$\mu$m; red).
Black circles show the positions and sizes of the S207 cluster, the S208
cluster, and the S208 association. The cyan crosses show the locations
of IRAS point sources, while the black crosses show other possible
star-forming regions, 2MASX J04183258+5326027 and Waterloo 1 (Wat 1).
White dashed eclipse shows the location of the bubble, which is
 identified by eye.}
\label{fig:WISEb134_S207S208}
\end{center}
\end{figure}

Figure~\ref{fig:WISEb134_S207S208} shows a {\it WISE} MIR pseudocolor
image around S207 and S208 with a wide field of view ($2^\circ \times
2^\circ$)\footnote{We used Image Co-addition with Optional Resolution
Enhancement software \citep{Masci2009} for producing custom-sized images
with {\it WISE} single-exposure images.}
centered at $(l, b) = (151.19^\circ, +2.13^\circ)$ in the Galactic 
coordinates.
Black circles show the positions and sizes of the S207 cluster, the S208 
cluster, and the S208 association (see Section~\ref{sec:S208cluster}).
The cyan crosses show the locations of IRAS point sources, while the
black crosses show some specific features, 2MASX J04183258+5326027 and
Waterloo 1 (Wat 1), in the SIMBAD online database.
2MASX J04183258+5326027 is an H$_2$O maser source, which is
a signature of massive star formation, while Wat 1 is a star cluster
\citep{Moffat1979}.
In the figure, a bubble-like structure (a white dashed ellipse) with
$\sim$22$\times$27\,arcmin radius is seen, where S207 appears to be
located around the edge.
The corresponding size of the bubble is $\sim$26$\times$33\,pc at the
distance of 4\,kpc, which is a relatively large structure compared to
those in the {\it WISE} MIR survey of massive star-forming regions by
\citet{Koenig2012}, 3--24\,pc, although it is smaller than superbubbles
with radius of $\sim$100--1000\,pc (e.g., \citealt{Sato2008}).
Bubbles are thought to be one of the major triggers of star formation
(e.g., \citealt{Deharveng2010}, \citealt{Koenig2012}).
Possible star-forming regions are located around the bubble: many IRAS
point sources are located inside and around the edge bubble, while 2MASX
J04183258+5326027 appears to be located around the edge of the bubble.
They suggest that the bubble triggers star formation in S207 and the
above regions. 
The bubble feature is not significant in the higher latitudes, while the
feature is prominent in the lower latitudes. 
Also, many possible star-forming regions are distributed in the lower
latitude on the line extending to S207, S208 (the S208 cluster and the
S208 association), Wat 1, and many IRAS point sources.
Because the radial velocity of Wat 1 is comparable to those of S207 and
S208, Wat 1 is suggested to actually be located at the identical
distance \citep{Foster2015}.
The positional relationship between S207, S208, and possible
star-forming regions suggests that the bubble also triggers star
formation in the lower latitudes and that sequential star formation is
occurring.
This is consistent with the age difference between the S207 cluster and
the S208 cluster, i.e., younger in the S207 and older in S208.
Because no massive stars\footnote{There is only one O-type star in the
field of view from the SIMBAD online data base, GSC 03719-00546, which
is located around the center of S207 and is a probable dominantly
exciting source of it.},
\ion{H}{2} regions, or supernova remnants are identified around the
center of the bubble in the SIMBAD online database, we cannot tell how
this bubble was made at this stage.

There are two major paradigms for triggered star formation: ``collect
and collapse'' \citep{Elmegreen1977} and ``radiation-driven implosion''
\citep{Bertoldi1989}.
There are no significant shell-like structures with neutral gas, which
are characteristics of the collect and collapse scenario
\citep{Deharveng2010}, around S207 and S208 in the \ion{H}{1} data (see
the bottom left panel of Figure~3 in \citealt{Foster2015}).
The pillar structures, which are often seen in locations where star
formation may be triggered via radiatively driven implosion (e.g.,
\citealt{Sugitani1994}), are seen around the bubble and in the lower
Galactic latitude of Wat 1.
They suggest that the star formation in this area is likely due to
radiation-driven implosion rather than the collect and collapse
scenario. 
However, further studies in larger scales by multiwavelength
observations (e.g., \citealt{Kobayashi2008}) are necessary in the future
for concluding the star formation triggers for the twin \ion{H}{2}
regions, S207 and S208.

\vspace{2em}

\acknowledgments

This work was supported by JSPS KAKENHI Grant Number 26800094 and MEXT
KAKENHI Grant Number 23103004. 
We thank the Subaru support staff, in particular, the MOIRCS support
astronomer Ichi Tanaka. We also thank Chihiro Tokoku for helpful
discussions on the observation.



\begin{table*}[h]
\caption{Properties of S208.}\label{tab:targets} 
\begin{center}
\begin{tabular}{llcccccccc}
\hline
\hline
Name & Sh 2-208 \\
Galactic longitude (deg) &  151.2870 (1) \\
Galactic latitude (deg) &  $+$1.9682 (1) \\
R.A. (J2000.0) & 04 19 32.92 (1) \\
Decl. (J2000.0) & $+$52 58 41.6 (1) \\
Photometric heliocentric distance (kpc) 
 & 7.6 (2), 9.4 (3), 10.0 (4) \\ 
Adopted photometric heliocentric distance (kpc) & 9 \\
Photometric Galactocentric distance$^{\rm a}$ (kpc) & $\simeq$16.5\\

Kinematic heliocentric distance (kpc) &  4.0 (5), 4.1 (6), 4.4 (7)\\
Adopted kinematic heliocentric distance (kpc) &  4 \\
Kinematic Galactocentric distance$^{\rm a}$ (kpc) &  $\simeq$12\\ 
Oxygen abundance $12 + \log {\rm (O/H)}$ & 7.91 (6, 8) \\
Metallicity [O/H] (dex)$^{\rm b}$ & $-$0.8 \\
\hline
\end{tabular}
\end{center}
{{\small {\bf Notes.} References are shown in the parenthesis. \\
$^{\rm a}$Assuming the solar Galactocentric distance $R_\odot =
 8.0$\,kpc. \\
$^{\rm b}$Assuming the solar abundance of $12+ \log {\rm (O/H)} = 8.73$
 \citep{Asplund2009}. \\
{\small \bf References. }
(1) SIMBAD {\citep{Wenger2000}}, 
(2) \citet{Moffat1979}, 
(3) \citet{Chini1984}, 
(4) \citet{Lahulla1985},
(5) \citet{Wouterloot1989}, 
(6) \citet{Caplan2000}, 
(7) \citet{Foster2015}, 
and (8) \citet{Rudolph2006}.}}
\end{table*}

\begin{table*}[!h]
\caption{Summary of MOIRCS Observations.} \label{tab:LOG}
\begin{center}
\begin{tabular}{lcccccccc}
\hline
\hline
Modes & Date & Band & $t_{\rm total}$ & $t$ & Coadd & $N_{\rm total}$
 & Seeing & Sky Condition\\ 
& & & (1) & (2) & (3) & (4) & & (5)\\
\hline  
$J$-long & 2008 Jan 14 & $J$ & 720 (480) & 120 & 1 & 6 (4) & $0''.6$& C \\

$H$-long & 2007 Nov 23 & $H$  & 420 (315) & 15 & 7 & 4 (3) & $0''.9$ & P \\

$K_S$-long & 2007 Nov 23 & $K_S$ & 960 (720) & 30 & 4 & 8 (6) & $1''.0$ & P \\

$J$-short & 2006 Nov 8 & $J$  & 52 (39) & 13 & 1 & 4 (3) & $1''.2$ &H \\
$H$-short & 2006 Nov 8 & $H$  & 52 (39) & 13 & 1 & 4 (3) & $1''.2$ &H\\
$K_S$-short & 2006 Nov 8 & $K_S$  & 52 (39) & 13 & 1 & 4 (3) & $1''.2$ & H\\
\hline
\end{tabular}
\end{center}
{{\small {\bf Notes.} 
Col. (1): total exposure time (s). The values for the sky frames are
 shown in parentheses.
Col. (2): single-exposure time (s).
Col. (3): number of coadd. 
Col. (4): total number of frames. 
Col.(5): C: cirrus, P: photometric, and H: high humidity. 
The values for the sky frames are shown in parentheses.}}
\end{table*}


\end{document}